\documentstyle[12pt,epsfig,cite,a4]{article}
\newcommand{\costhe} {\cos\theta}
\newcommand{\acosthe}   {|\costhe\,|}

\def\etal{\textit{et al.}}
\def\dEdx{\ifmmode {{\mathrm d}E/{\mathrm d}x} \else
                   {${\mathrm d}E/{\mathrm d}x$} \fi }
\def\DEdx{\ifmmode {{\mathrm d}E/{\mathrm d}x} \else
                   {${\mathrm d}E/{\mathrm d}x$} \fi }
\def\dedx{\mbox{${\mathrm d}E / {\mathrm d}x$}}
\def\sigmadedx{\mbox{$\sigma({\mathrm d} E / {\mathrm d} x)$}}
\def\ndedx{\mbox{$N_{{\mathrm d} E / {\mathrm d} x}$}}

\def\p{\mbox{$p$}}

\def\kapi{\mbox{${\mathrm K}\pi$}}

\def\ksks{\mbox{${\mathrm K_s^0}{\mathrm K_s^0}$}}
\def\kaka{\mbox{${\mathrm K}{\mathrm K}$}}
\def\kakalik{\mbox{${\mathrm K^{\pm}}{\mathrm K^{\pm}}$}}

\def\kzerokzero{\mbox{${\mathrm K^0}{\mathrm K^0}$}}
\def\antikzeroantikzero{\mbox{${\overline{\mathrm K^0}}\thinspace{\overline{\mathrm K^0}}$}}
\def\kzeroantikzero{\mbox{${\mathrm K^0}{\mathrm {\overline {K^0}}}$}}
\def\pipi{\mbox{${\mathrm \pi\pi}$}}
\def\pipilik{\mbox{${\mathrm \pi^{\pm}\pi^{\pm}}$}}

\def\kaprot{\mbox{${\mathrm Kp}$}}

\def\protprot{\mbox{${\mathrm p}{\mathrm p}$}}
\def\pion{\mbox{${\mathrm \pi}$}}

\def\ka{\mbox{${\mathrm K}$}}

\def\elecm{\mbox{${\mathrm e^-}$}}

\def\gevc{\mbox{GeV/$c$}}
\def\gev{\mbox{GeV}}
\def\gevinvone{\mbox{${\mathrm GeV^{-1}}$}}
\def\gevinvtwo{\mbox{${\mathrm GeV^{-2}}$}}
\def\mev{\mbox{MeV}}
\def\mevc{\mbox{MeV/$c$}}

\def\q{\mbox{$Q$}}

\def\qlt{\mbox{$Q<\thinspace$}}
\def\qlteight{\mbox{$Q<\thinspace0.8\thinspace\gev$}}
\def\qltsix{\mbox{$Q<\thinspace0.6\thinspace\gev$}}
\def\qltfive{\mbox{$Q<\thinspace0.05\thinspace\gev$}}
\def\qlttwo{\mbox{$Q<\thinspace2.0\thinspace\gev$}}
\def\qnorm{\mbox{$0.6\thinspace<Q<\thinspace2.0\thinspace\gev$}}
\def\qnormsys{\mbox{$0.8\thinspace<Q<\thinspace2.0\thinspace\gev$}}

\def\nor{\mbox{$N$}}
\def\lamb{\mbox{$\lambda$}}
\def\Rad{\mbox{$R$}}
\def\Radius{\mbox{$R_0$}}

\def\Zzero{\mbox{$\mathrm{Z}^0$}}
\def\zzero{\mbox{$\mathrm{Z}^0$}}
\def\lambval{\mbox{0.82}}
\def\radval{\mbox{0.56}}
\def\lambstat{\mbox{0.22}}
\def\radstat{\mbox{0.08}}
\def\lambsysa{~0.17}
\def\lambsysb{~0.12}
\def\radsysa{~0.08}
\def\radsysb{~0.06}
\def\lambplain{\mbox{\lambval~$\pm$~\lambstat}}
\def\radplain{\mbox{\radval~$\pm$~\radstat}}
\def\lambresult{\mbox{$\lambplain~^{+\lambsysa}_{-\lambsysb}$}}
\def\radresult{\mbox{$\radplain~^{+\radsysa}_{-\radsysb}$}}
\def\lambresulttable{\mbox{$\lambplain~^{+\lambsysa}_{-\lambsysb}$}}
\def\radresulttable{\mbox{$\radplain~^{+\radsysa}_{-\radsysb}$}}
\begin{document}
\begin{titlepage}
\begin{center}
{\Large EUROPEAN LABORATORY FOR PARTICLE PHYSICS}
\end{center}
\bigskip
\begin{flushright}
{\large CERN-EP/99-163\\
\today}
\end{flushright}
\vfill
\begin{center}
{\LARGE\bf  
 Bose-Einstein correlations in \kakalik\ pairs from \zzero\ decays into
 two hadronic jets} \\
\end{center}
\bigskip
\begin{center}
{\LARGE The OPAL collaboration}
\end{center}
\bigskip
\begin{abstract}
\noindent
Bose-Einstein correlations in pairs of charged kaons  
produced in a sample of 3.9 million hadronic \zzero\ 
decays have been measured with the OPAL experiment at LEP. Charged 
kaons were identified in the central tracking 
detector using their specific energy loss in the drift chamber gas. 
The correlation function was studied in two-jet events using 
a double ratio, formed by the number of like-sign pairs normalised 
by a reference sample in the data,
divided by the same ratio in a Monte Carlo simulation. The enhancement 
at small values of the four-momentum difference of the pair was parametrised 
using a Gaussian function. The parameters of the Bose-Einstein
correlations were measured to be \mbox{\lamb\ = \lambresult } for 
the strength and \mbox{\Radius\ = \radresult\ fm} for the kaon source
radius, where the first errors are statistical and the second
systematic. Corrections for final-state interactions are discussed.
\end{abstract}
\vfill

\vfill

\begin{center}
{\large (Submitted to European Physical Journal C)}
\end{center}

\end{titlepage}


\begin{center}{\Large        The OPAL Collaboration
}\end{center}
\begin{center}{
G.\thinspace Abbiendi$^{  2}$,
K.\thinspace Ackerstaff$^{  8}$,
P.F.\thinspace Akesson$^{  3}$,
G.\thinspace Alexander$^{ 23}$,
J.\thinspace Allison$^{ 16}$,
K.J.\thinspace Anderson$^{  9}$,
S.\thinspace Arcelli$^{ 17}$,
S.\thinspace Asai$^{ 24}$,
S.F.\thinspace Ashby$^{  1}$,
D.\thinspace Axen$^{ 29}$
% ...
}\end{center}\bigskip
$^{  1}$School of Physics and Astronomy, University of Birmingham,
Birmingham B15 2TT, UK
\newline
$^{  2}$Dipartimento di Fisica dell' Universit\`a di Bologna and INFN,
I-40126 Bologna, Italy
\newline
$^{  3}$Physikalisches Institut, Universit\"at Bonn,
D-53115 Bonn, Germany
\newline
% ...
$^{  8}$CERN, European Organisation for Particle Physics,
CH-1211 Geneva 23, Switzerland
\newline
$^{  9}$Enrico Fermi Institute and Department of Physics,
University of Chicago, Chicago IL 60637, USA
\newline
% ...
$^{ 16}$Department of Physics, Schuster Laboratory, The University,
Manchester M13 9PL, UK
\newline
$^{ 17}$Department of Physics, University of Maryland,
College Park, MD 20742, USA
\newline
% ...
$^{ 23}$Department of Physics and Astronomy, Tel Aviv University,
Tel Aviv 69978, Israel
\newline
$^{ 24}$International Centre for Elementary Particle Physics and
Department of Physics, University of Tokyo, Tokyo 113-0033, and
Kobe University, Kobe 657-8501, Japan
\newline
% ...
$^{ 29}$University of British Columbia, Department of Physics,
Vancouver BC V6T 1Z1, Canada
\newline
% ...

\section{Introduction} \label{s:1}
Intensity interferometry was applied in 1953 by Hanbury-Brown and
Twiss~\cite{HBT} in radio astronomy in order to estimate the 
spatial extension
of stars (HBT effect). In particle reactions which lead to multi-hadronic final
states the HBT effect manifests itself as a constructive interference
between two identical bosons --- the so-called Bose-Einstein (BE)
correlation, which is now well known and was first observed by Goldhaber
 et al.~\cite{Goldhaber1} in $\bar {\mathrm p}$p annihilations. 
There is interest
in the quantum mechanical aspects of the BE correlations, but
they are also used to
estimate the dimensions of the source of the identical bosons. 
BE correlation
studies of pion pairs have been carried out in a large variety of
particle interactions and over a wide range of 
energies~\cite{Goldhaber2,Haywood,Dewolf}. Recently, pion BE correlations
have been investigated in
connection with the ${\mathrm W}$ mass measurement in 
${\mathrm e}^+ {\mathrm e}^- \to {\mathrm W}^+ {\mathrm W}^-$
reactions at LEP at centre-of-mass energies above 161 GeV~\cite{pipiwwpaperOPAL,pipiwwpaperDELPHI}.

Compared to the abundant information now available on 
BE correlations in pion pairs, knowledge of the correlations in 
identical strange
boson pairs is scarce and, until recently, was mainly limited to the
\ksks\ system. \ksks\ pairs may exhibit a BE correlation
enhancement near threshold even if the origin is not a 
\kzerokzero\ or \antikzeroantikzero\ system but a \kzeroantikzero\ 
boson-antiboson system. A \ksks\ low-mass enhancement has recently
been observed in hadronic \zzero\ 
decays~\cite{kskspaperOPAL,kskspaperALEPH,kskspaperDELPHI,kkpaperDELPHI}.
However, the
interpretation of this enhancement 
as a pure BE correlation effect is complicated by 
the possible presence of the f$_0(980)$ decay into
\kzeroantikzero . Recently it has been pointed
out that the information from BE correlation
studies of the \kakalik\ system, 
which cannot result from the f$_0(980)$ decay, can serve as 
an effective tool in setting a limit on the resonant 
fraction of the \ksks\ BE enhancement~\cite{preprintalex_lipkin}.

The study of \kakalik\ BE
correlations also has a bearing on the relation between the dimension of
the emitting source and the mass of the emitted particles; from recent 
measurements it has been pointed out that the source dimension seems to decrease as 
the mass increases~\cite{preprintGideon}. Several models have 
been proposed to account for this behaviour~\cite{preprintGideon,preprintBialas}.

This paper presents a study
of \kakalik\ BE correlations using the high statistics
sample of \zzero\ hadronic events recorded by the OPAL detector at
LEP. The paper is organised as follows. In Section~\ref{s:2} the 
methodology used for measuring the BE correlations is
presented. The event and track selections are described in Section~\ref{s:3}. In Section~\ref{s:4} the analysis of the data is presented,
and in Section~\ref{s:5} the systematic effects are studied. 
Finally, the conclusions are drawn in Section~\ref{s:6}.

\section{Analysis method}\label{s:2}
The BE correlation function for two identical bosons is defined as:
\begin{eqnarray} \label{eq:Cp1p2}
 C(p_1,p_2)= \frac{\rho(p_1,p_2)}{\rho(p_1)\rho(p_2)}{\mathrm ,}
\end{eqnarray} 
where $\rho(p_1,p_2)=(1/\sigma)({\mathrm d}^8\sigma/{\mathrm d}^4
\p_1{\mathrm d}^4 \p_2)$ is the two-particle phase space density subject to BE
symmetry, and $\rho(p)$ is the corresponding single particle quantity for a
particle with four-momentum \p . The correlation function can be
studied as a function of the four-momentum difference of the pair, \q
, where $Q^2$ = \mbox{$-(p_1-p_2)^2$} = \mbox{$M^2 - 4m^2_{{\mathrm
boson}}$}, and $M$ is the invariant mass of the pair of bosons 
each of mass $m_{{\mathrm boson}}$.

From the study of the correlation in pairs one can 
determine the geometrical and dynamical properties of the 
emitting source. For a static sphere of emitters 
with Gaussian density, the correlation function is parametrised with 
the Goldhaber function~\cite{Goldhaber2} as:
\begin{eqnarray} \label{eq:Cgauss}
 C(Q) = 1 + \lambda\thinspace e^{-(RQ)^2}{\mathrm ,}
\end{eqnarray} 
where \lamb, the strength of the correlation, is 0
for a completely coherent source and 1 for a completely incoherent
one. The parameter \Rad\ in \gevinvone\ is related to the radius of the 
source, \Radius , through the relation $\Radius=\Rad\hbar c$.
 
The two-particle phase space density $\rho(p_1,p_2)$ is obtained from a
sample of pairs of identical bosons. In this analysis this 
sample, called the like-sign
sample, is formed by pairs of kaons with the same charge. The 
denominator of Equation~\ref{eq:Cp1p2} $\rho(p_1)\rho(p_2)$  
is in practice replaced by a reference 
distribution $\rho_0(p_1,p_2)$, which resembles
$\rho(p_1,p_2)$ in all aspects except in the BE
symmetry. A perfect reference sample should have the following properties: 
absence of BE correlations, and presence of the same correlations 
as in the the like-sign sample, arising from    
energy-momentum and charge conservation, 
the topology and global properties of the events, and resonance or
long-lived particle decays. 

The principal difficulty for measurements of BE correlations is in the  
definition of the reference sample from which $\rho_0(p_1,p_2)$ is obtained.
When correlations amongst like-sign boson pairs are measured, the
obvious reference sample is provided by unlike-sign boson pairs. 
Unfortunately, the \q\ distribution of the unlike-sign pairs 
includes prominent peaks due to neutral meson resonances \mbox{(e.g. $\phi
\rightarrow {\mathrm K^{+}K^{-}}$)}. An alternative 
reference sample can be derived from Monte Carlo simulations in which
BE correlations are not included. This relies on a
correct simulation of the physics in the complete absence of any BE
effect, and a correct modeling of detector effects.
A third type of reference sample can be obtained using the methods of
event- or hemisphere-mixing, where particles from different events or 
hemispheres are combined. The last, least model-dependent method 
was used in the present work.

Each event was divided into two hemispheres separated by 
the plane perpendicular to the thrust axis and containing the 
interaction point. A charged kaon track in one hemisphere 
was combined with a kaon track of the same charge found in 
the opposite hemisphere
after reflecting the momentum of this second track through the origin. To
ensure that the like-sign and reference samples were independent, the 
two kaon tracks forming a pair in the like-sign sample were required to be
in the same hemisphere. If all the events have two back-to-back
jets, the reference sample of hemisphere-mixed pairs will be similar
to the like-sign pair sample apart from the lack of BE correlations.
Monte Carlo studies have shown that the hemisphere-mixing technique only works 
in symmetric topologies such as back-to-back
jets. Therefore, two-jet events were selected by requiring 
a high value of the thrust. 
The event- and hemisphere-mixing techniques  
have already been used in other experiments~\cite{pipieebarpaperALEPH,pipieebarpaperDELPHI}. 


\section{Event and track selection} \label{s:3}
A detailed description of the OPAL detector can be found in 
reference~\cite{opaldetector2}. This analysis is primarily 
based on information from the central tracking detectors, consisting 
of a silicon microvertex
detector, a vertex drift chamber, a jet chamber and $z$-chambers\footnote{A
right handed coordinate system is used, with positive $z$ along
the \elecm\ beam direction and $x$ pointing towards the centre of the
LEP ring. The polar and azimuthal angles are denoted by $\theta$
and $\phi$, and the
origin is taken to be the centre of the detector.}, 
all of which lie within an axial
magnetic field of 0.435~T. The jet
chamber, which has an outer radius of 185~cm, provides up to 159
measurements of space points per track with a resolution in the 
\mbox{$r$--$\phi$} plane of about 135~$\mu$m and a
transverse momentum\footnote{The projection onto the plane
perpendicular to the beam axis.}
resolution of $\sigma_{p_T}/p_{T}$ = {\mbox{$\sqrt{(0.02)^2 + (0.0015\thinspace p_T)^2}$},
with $p_T$ in \gevc . 
Particle identification in the jet chamber is possible using 
the measurement of the specific energy loss \dedx\ \cite{dedxpaperold}
with a resolution of approximately of $3\%$ for  
high-momentum tracks in hadronic decays~\cite{dedxpaper}.
Since the identification of charged kaons using \dedx\ is crucial to
this analysis, the calibration of the energy loss over the many years
of data taking was checked and improved when necessary.
Control samples of particles identified by techniques other than the
energy loss were used to remove year-to-year 
variations in the measured \dedx\ and to resolve discrepancies 
between the measured \dedx\ in the data and the
theoretical \dedx\ in the Monte Carlo.

This analysis used a sample of about 3.9 million hadronic \Zzero\
decays recorded at LEP between the years 1992 and 1995. 
A sample of 6.75 million 
Monte Carlo hadronic events generated with JETSET 7.4~\cite{jetset}
and tuned to reproduce the global features of the events~\cite{opaltune}
was also used. The generated events were processed 
through a detailed simulation of 
the experiment~\cite{opalsim} and subjected to 
the same event and track selection as the data. A detailed
description of the selection of hadronic events is 
given in~\cite{opalhadronic}. Events with two clear back-to-back
jets, necessary for the 
proper functioning of the hemisphere-mixing technique,
were selected by requiring that the thrust value was larger than 0.95. 
About 30$\%$ of the events passed the thrust cut. 
 
Charged tracks were required to have a minimum transverse momentum 
of 150~\mevc , a maximum reconstructed 
momentum of 60~\gevc , a distance of closest 
approach to the interaction point less than 0.5~cm
in the plane orthogonal to the beam direction and the corresponding
distance along the beam direction of less than 40~cm. The first measured
point had to be within a radius of 75~cm from the interaction vertex. The 
cuts on the transverse and longitudinal 
distances of closest approach help to remove
particles from long-lived decays. About 50$\%$ of the reconstructed tracks 
passed these selections. 

Kaons were identified using the \dedx\ measurement. Only tracks with
at least 20 hits available for the measurement of \dedx\ and with 
a polar angle satisfying $\acosthe\thinspace<\thinspace0.9$ 
were considered. For each
track, a $\chi^2$ probability was formed for each stable particle
hypothesis: e, $\mu$, \pion , \ka\ and p. A track was identified as a
kaon candidate if it had a probability of at least 10$\%$ 
of being a kaon and if the probability of being a kaon was larger than the 
probability of being any other of the above particle species. In addition, in
order to reject pions, each track was
required to have a pion probability less than 5$\%$. 
Electrons from photon conversions were rejected using a neural network
algorithm as described in~\cite{Rbpaper}. According to the Monte
Carlo, approximately 34$\%$ of the 
true kaons passed these requirements and the estimated kaon purity of 
the track sample was about 72$\%$ on average, with 
variations between 50$\%$ and 97$\%$ depending on 
the momentum of the track. The lowest purity 
corresponded to the momentum range between 0.9 and 
1.5 \gevc\ where the pion and kaon bands overlap in \dedx ~\cite{dedxpaper}. 
In the selected kaon track sample the fraction of pions was estimated
as 17$\%$, that of protons as 11$\%$, and the contribution of muons
and electrons was negligible.

\section{Data analysis}\label{s:4}
The \q\ distributions of the like-sign kaon candidate pairs and the 
hemisphere-mixed kaon candidate pairs were determined using 
tracks that passed the selections described in
Section~\ref{s:3}. In the data, 76063 like-sign and 
98558 hemisphere-mixed kaon candidate pairs with \qlttwo\ were
selected. In the Monte Carlo, the corresponding numbers of 
kaon candidate pairs were 109601 and 136785 respectively.
 
Since the BE correlation manifests itself only for identical particles, 
it was necessary to correct for impurities. The Monte Carlo 
predicted that the main contamination
in the sample of like-sign kaon candidate pairs was from \kapi\ and
\kaprot\ pairs. In this sample, and for pairs with \qlttwo , the
estimated fraction of \kaka\ pairs was about 48$\%$, the \kapi\
fraction about 27$\%$
and the \kaprot\ fraction about 13$\%$. The contamination from
pairs susceptible to BE correlations or Fermi-Dirac 
anti-correlations (i.e. \pipi\
and \protprot ) was negligible, of the order of 3$\%$ for \pipi\ and less than
1$\%$ for \protprot\ pairs. The fraction 
of \kaka\ pairs was constant over the whole \q\ range.
The sample of hemisphere-mixed kaon candidate pairs had 
an almost identical composition.

The Monte Carlo non-\kaka\ \q\ distribution was subtracted from the 
data distribution using the fraction given by the simulation and 
normalised to the number of data pairs. Both like-sign and
hemisphere-mixed pair distributions were 
corrected using this technique. Figure~\ref{f:01} shows 
these corrected \q\ distributions for the data and 
the Monte Carlo in which BE correlations were not simulated. 
The hemisphere-mixed \q\ distributions were normalised to the like-sign 
\q\ distributions in the region \qnorm\ where no BE correlations are expected. 
Both data and Monte Carlo distributions show a similar behaviour at high values
of \q : the hemisphere-mixed distribution is above the like-sign one
in the region $0.7\thinspace<\q<\thinspace1.2\thinspace\gev$; there is a 
cross-over of both distributions at \q\ about 1.2 \gev ; and the like-sign
distribution is above the hemisphere-mixed one at $\q>1.2~\gev$. 

Figure~\ref{f:02} shows 
the ratio $N_{++}(Q)/N_{mix}(Q)$ in both the data and the
Monte Carlo, where $N_{++}(Q)$ is 
the number of like-sign pairs and $N_{mix}(Q)$ is the number 
of hemisphere-mixed pairs as functions of $Q$. The first
two bins of the distribution shown in the figure were combined 
due to the small statistics. 
The data distribution shows a clear 
enhancement in the region \qlt\ 0.3 \gev. There is also a rise
of the correlation at high values of \q\ normally attributed to long-range 
correlations. Indeed, the slope of the correlations at high \q\ is 
well reproduced by the Monte Carlo. The ratio $N_{++}(Q)/N_{mix}(Q)$  
in the Monte Carlo deviates slightly from unity at high \q\ and falls slowly 
with decreasing \q , probably due to features of string fragmentation and
local conservation of charge and strangeness.
These effects can be taken into account if both like-sign 
and hemisphere-mixed data distributions are divided by the corresponding 
Monte Carlo distributions. 

The correlation function was therefore defined as the double-ratio:
\begin{eqnarray} \label{eq:Cmix}
 C_{mix}(Q) =
\frac{N_{++}^{data}(Q)}{N_{mix}^{data}(Q)}/
\frac{N_{++}^{MC}(Q)}{N_{mix}^{MC}(Q)} {\mathrm ,}
\end{eqnarray} 
and was parametrised using a modified version of Equation~\ref{eq:Cgauss}:
\begin{eqnarray} \label{eq:Cfinal}
 C(Q) = N\thinspace(1 + \lambda\thinspace e^{-(RQ)^2})\thinspace
(1+\delta\thinspace Q+\epsilon\thinspace Q^2) {\mathrm ,}
\end{eqnarray}
where \nor\ is a normalisation factor, and 
the empirical term $(1+\delta\thinspace Q+\epsilon\thinspace Q^2)$ 
accounts for the behaviour of the correlation function at high $Q$
values due to any remaining long-range correlations.
Figure~\ref{f:03} shows the correlation $C_{mix}(Q)$
with the result of the fit. The fitted parameters and the correlation 
coefficients between them are given in Table~\ref{t:01}.
The fit has a $\chi^2$ = 43 for 35 degrees of freedom. 
\begin{table}[h]
\centering
{\begin{tabular}{|l|r @{.} l||ccccc|}
\hline
\hline
\multicolumn{3}{|l||}{} & \multicolumn{5}{|c|}{Correlation coefficients}\\
\hline
       &\multicolumn{2}{c||}{fitted value}    &\nor\ & \lamb\ &\Radius\ & $\delta$ & $\epsilon$ \\
\hline
\nor\         &0&97 $\pm$ 0.11 &  & $-0.59$ & $+0.52$ & $-0.96$ & $+0.85$  \\
\lamb\        &0&82 $\pm$ 0.22 &  &         & $+0.17$ & $+0.58$ & $-0.52$ \\
\Radius\ (fm) &0&56 $\pm$ 0.08 &     &    &           & $-0.42$  & $+0.29$  \\
$\delta$ (\gevinvone )  &$-$0&07 $\pm$ 0.16 & & &  &               & $-0.96$ \\
$\epsilon$ (\gevinvtwo )&0&06 $\pm$ 0.06 & & &  &   &     \\
\hline 
\hline 
\end{tabular}}
\caption{\label{t:01} Fitted parameters 
and correlation coefficients obtained 
in the Gaussian parametrisation of the correlation function
$C_{mix}(Q)$. The uncertainties on the parameters 
are statistical only.}  
\end{table}

\section{Systematic effects} \label{s:5}
Systematic effects arising from the event and track 
selections, the modeling of \dedx\ in the Monte Carlo, 
the parametrisation of the correlation function and 
the choice of the reference sample are considered. 
In each case, the result of the fit to the correlation
function is given in Table~\ref{t:02}, with row~(a) giving the result
of the basic fit discussed in the previous section.

The overall systematic uncertainties in the 
parameters \lamb\ and \Radius\ were calculated as the largest 
single deviations between the parameters of the fits from rows~(b) to (l), 
and the parameters of the basic fit in row~(a). 
The final values of the parameters are
\begin{eqnarray} 
 \lamb  & = & \lambresult    \nonumber  \\
 \Radius & = & \radresult ~{\mathrm fm}. \nonumber
\end{eqnarray}

\noindent 
The mismodeling of the kaon momentum spectrum, residual BE
correlations in the reference sample, the origins of the kaons and
final-state interaction corrections are also discussed in this section.

\subsection{Event and track selection}\label{s:5.1}
\begin{itemize}
\item 
As discussed in Section~\ref{s:2}, the hemisphere-mixing technique
only works when two-jet events are selected.
The analysis was repeated for events selected using a cone jet 
finding algorithm~\cite{conejetpaper} instead 
of the cut in thrust (row~(b)). 
The value of the thrust cut was also changed 
from 0.95 to 0.93 and the analysis repeated (row~(c)). 
\item
To obtain a purer sample of kaons, tracks with momenta in the 
pion-kaon \dedx\ overlap 
region, $0.9\thinspace<p<\thinspace1.5$~\gevc , 
were rejected (row~(d)). 
\item
The minimum number of \dedx\ hits required for each track was increased from
20 to 40 (row~(e)). 
\end{itemize} 
\subsection{Parametrisation of \dedx }\label{s:5.2}
Since the analysis relies to a large extent on the separation of pions 
from kaons, it is especially important to understand 
the \dedx\ measurements of the copiously-produced pions. A sample
of pions was identified in ${\mathrm K^0_s} \rightarrow \pi^{+}\pi^{-}$ 
decays~\cite{kskspaperOPAL} and was used to estimate the mismodeling of the normalised 
ionisation energy loss $N_{\dedx}^{\sigma}$~\cite{Rbpaper}. 
The normalised \dedx\ is defined as 
$N_{\dedx}^{\sigma} = (\dedx - (\dedx)_0)/(\sigmadedx)_0$. Here,
$\dedx$ is the measured value, while $(\dedx)_0$ 
and $(\sigmadedx)_0$ are the expected value and the expected error
assuming the track to be a pion. 
This analysis showed that the mean value and the width of 
the normalised \dedx\ distribution were both known 
to within $\pm10\%$ of $\sigma$. By changing in the 
Monte Carlo the normalised \dedx\ of
both kaons and pions by their known uncertainties, 
the fraction of non-\kaka\ pairs 
was found to vary by up to $\pm10\%$.
The BE analysis was consequently repeated with the assumed 
impurity value set to 57$\%$ and 47$\%$ (rows~(f) and (f$'$)).

\subsection{Fit of the correlation function}\label{s:5.3}
\begin{itemize}
\item
The binning of the \q\ distributions was modified from 50 \mev\ to
20 \mev\ (row~(g)). 
\item
The \q\ distribution normalisation range was changed from \qnorm\ to
\qnormsys\ (row~(h)).
\item
The fit was repeated with the first bin \qltfive , excluded. This was
done to test the effect of 
potential problems at low \q\ because of the limited 
resolution~\cite{pipieebarpaperOPAL} (row~(i)).
\item
If the use of the double-ratio removes all 
correlations other than BE, such as long-range correlations at high \q , then
the empirical term \mbox{$(1+\delta\thinspace Q+\epsilon\thinspace
Q^2)$} of Equation~\ref{eq:Cfinal} would not be
necessary. $C_{mix}(Q)$ was parametrised using the simplified function:
\begin{eqnarray} \label{eq:Cgauss3param}
 C(Q) = N\thinspace(1 + \lambda\thinspace e^{-(RQ)^2}). 
\end{eqnarray}
The parameters of the fit are given in row~(j). The fit 
has $\chi^2$ = 49 for 37 degrees of freedom .
This fit was also repeated with the range limited to $\q<1.5$~\gev\ 
to reduce possible long-range correlations (row~(k)).
\end{itemize}
\subsection{The reference sample}\label{s:5.4}
The BE correlation was also measured using Monte Carlo like-sign pairs 
as the reference sample in the sub-sample of two-jet events (row~(l)). 
The correlation function in this case was defined as: 
\begin{eqnarray} \label{eq:Cmc1}
 C_{MC}(Q) = \frac{N_{++}^{data}(Q)}{N_{++}^{MC}(Q)} {\mathrm .}
\end{eqnarray} 
Figure~\ref{f:04} shows the correlation
function $C_{MC}(Q)$ together with 
the results of a fit using Equation~\ref{eq:Cfinal}. The values
of the parameters are: \nor\ = 0.88 $\pm$ 0.05;
\lamb\ = 0.92 $\pm$ 0.17; \mbox{\Radius\ = 0.59 $\pm$ 0.06 fm};
\mbox{$\delta$ = 0.04 $\pm$ 0.08 \gevinvone }; \mbox{$\epsilon$ = 0.05
$\pm$ 0.03 \gevinvtwo }, where the errors are statistical only. The
fit has $\chi^2$ = 42 for 35 degrees of freedom. 

Although there are known imperfections in the simulation, 
particularly at low momenta in the kaon momentum 
spectrum (see section~\ref{s:5.5}), these results were taken as an indication
of systematic differences between the hemisphere-mixing method and the
simple use of a Monte Carlo reference sample.
\begin{table}[h]
\centering
{\begin{tabular}{|l|c|c|c|}
\hline
\hline
 fit variation & \lamb\ & \Radius\ (fm)& $\chi^2$/DoF \\
\hline
 (a) basic fit        & \lambplain\  & \radplain\ &43/35\\
\hline 
 (b) cone jet finding  &0.90 $\pm$ 0.21  &0.58 $\pm$ 0.13 &45/35\\
\hline 
 (c) thrust $>$ 0.93  &0.88 $\pm$ 0.19  &0.53 $\pm$ 0.06 &42/35\\
\hline 
 (d) cut on \p\         &0.89 $\pm$ 0.23  &0.58 $\pm$ 0.08 &37/35\\
\hline 
 (e) minimum \ndedx\ = 40 &0.79 $\pm$ 0.27  &0.62 $\pm$ 0.13 &43/35\\
\hline 
 (f) \dedx\ $+10\%$           &0.99 $\pm$ 0.26 &0.55 $\pm$ 0.07 &44/35\\
\hline 
 (f$'$) \dedx\ $-10\%$       &0.70 $\pm$ 0.20 &0.58 $\pm$ 0.11 &42/35\\
\hline 
 (g) \q\ binning of 20 \mev\  &0.77 $\pm$ 0.22 &0.50 $\pm$ 0.06 &80/95\\
\hline 
 (h) normalisation \qnormsys\ &0.82 $\pm$ 0.21 &0.56 $\pm$ 0.08 &43/35\\
\hline 
 (i) \q\ lower limit 0.05 \gev\ &0.82 $\pm$ 0.23 &0.57 $\pm$ 0.09 &43/34\\
\hline 
 (j) fit using Equation~\ref{eq:Cgauss3param} &0.85 $\pm$ 0.21 &0.64 $\pm$ 0.08 &49/37\\
\hline 
 (k) \q\ upper limit 1.5 \gev\ &0.81 $\pm$ 0.21 &0.59 $\pm$ 0.08 &38/28\\
\hline 
 (l) Monte Carlo as reference sample&0.92 $\pm$ 0.17 &0.59 $\pm$ 0.06 &42/35\\
\hline 
                                  &$+$\lambsysa\ & $+$\radsysa\ & \\ 
 Overall systematic error         &$-$\lambsysb\ & $-$\radsysb\ & \\
\hline  
\hline 
\end{tabular}}
\caption{\label{t:02} Results of various fits of the BE correlation
function; the quoted errors associated with 
$\lambda$ and \Radius\ are statistical only. The results in
row~(a) correspond to the basic fit of Section~\ref{s:4} while the
other lines show the results obtained when modifying certain criteria as 
explained in the text. The overall systematic uncertainties in the 
parameters \lamb\ and \Radius\ were calculated as the largest 
single deviations between the parameters of the fits from rows~(b) to (l), 
and the parameters of the basic fit in row~(a). The final
systematic uncertainties in $\lambda$ and \Radius\ are given in the last row.}
\end{table}

\subsection{Mismodeling of the kaon momentum spectrum}\label{s:5.5}
If the simulation were perfect, one would expect to get 
the same results by measuring the correlation function with
$C_{mix}(Q)$ and with $C_{MC}(Q)$. The double-ratio
\begin{eqnarray} \label{eq:Cmc2}
 C_{mix}(Q) = 
\frac{N_{++}^{data}(Q)}{N_{mix}^{data}(Q)}/
\frac{N_{++}^{MC}(Q)}{N_{mix}^{MC}(Q)} \equiv 
\frac{N_{++}^{data}(Q)}{N_{++}^{MC}(Q)}/
\frac{N_{mix}^{data}(Q)}{N_{mix}^{MC}(Q)}  
\end{eqnarray} 
is equivalent to $C_{MC}(Q)$ only if the hemisphere-mixed sample is 
perfectly modelled, which 
implies \mbox{$N_{mix}^{data}(Q)/N_{mix}^{MC}(Q) = 1$}. 
Figure~\ref{f:05} shows this ratio; the general 
agreement is good, although at \qlteight\ there is some indication that
the ratio is below unity.

Any important mismodeling of the Monte
Carlo would indicate that the results obtained with $C_{MC}(Q)$ and 
with $C_{mix}(Q)$ (since the Monte Carlo is used for normalising this
correlation function) were not reliable. Therefore, the distribution 
of Monte Carlo tracks in a sample free from BE correlations 
was compared to the same distribution in the data. 
Figure~\ref{f:06} shows the momentum spectrum of kaon
candidate tracks of the hemisphere-mixed sample in the data and
in the Monte Carlo normalised to the same total 
number of pairs. At low momentum, the
Monte Carlo does not describe the data spectrum well and differences
are seen at the $\pm15\%$ level. Studies of the
differential cross-section of kaons in hadronic events in both the 
data and the Monte Carlo events, generated with JETSET 7.4 and tuned
according to OPAL data, showed that 
the simulation predicted a kaon spectrum which is too soft. 

As a check of the stability of the results obtained with
$C_{mix}(Q)$, the \q\ spectra of 
both like-sign and hemisphere-mixed pairs were reweighted. Each pair 
of kaons in the Monte Carlo was reweighted by the product of the weights of
each kaon in the pair, where the weight was 
obtained in bins of momentum by dividing the data momentum spectrum by
the Monte Carlo spectrum. The final measurement of the 
correlation function did not change significantly 
after reweighting, \lamb\ varied by $+$0.03 and \Radius\ by $-$0.01 fm. 
The same exercise was done to check the results obtained with
$C_{MC}(Q)$. In this case, \lamb\ varied more significantly, by $+$0.12, 
and \Radius\ by $-$0.01 fm. 
Thus, by the use of a double-ratio technique, the correlation 
function $C_{mix}(Q)$ was found to be less sensitive to the 
Monte Carlo mismodeling than the correlation function $C_{MC}(Q)$. 
\subsection{Residual BE correlations in the reference sample} \label{s:5.6}
As suggested in \cite{pipieebarpaperALEPH}, residual BE correlations
could be a source of imperfection in the reference sample. To check that
the hemisphere-mixed sample was free of effects induced by the BE correlations,
a Monte Carlo study was done using the generator JETSET 7.4. Hadronic 
events were generated with and without BE
correlations, with the BE correlations simulated assuming the Goldhaber
parametrisation described in Section~\ref{s:2}. The shape 
of the \q\ distribution
of the hemisphere-mixed pairs remained unchanged after
including the BE correlations in the generation. The 
ratio $N_{mix}^{with~BE}(Q)/N_{mix}^{without~BE}(Q)$ was consistent
with unity in the full range of \q, demonstrating that the 
reference sample was free from effects due to residual BE correlations.
\subsection{Sources of kaons} \label{s:5.7}
In a substantial fraction of the kaon pairs, one or more of the kaons result 
from a long-lived particle decay --- in such cases the kaon pairs cannot 
exhibit BE correlations. It is therefore useful to separate the various sources of 
kaons and to classify the parent particles to estimate the maximum
possible value of \lamb , as suggested in~\cite{Haywood}. 
On the other hand, some studies~\cite{Ellis,Wiedemann,Bolz} 
have suggested that the correlation function is narrowed by the 
contribution of decay products of long-lived sources, and also that 
resonance decays induce a pair-momentum dependence of the radius. 

The kaon sources as predicted by the JETSET Monte Carlo simulation 
are given in Table~\ref{t:03}. These have been
classified as in~\cite{Haywood,Ellis} into two main groups: 
long-lived sources with life-time $c\tau >$ 10 fm, and  
short-lived sources with life-time $c\tau <$ 10 fm.
The table shows that the fraction of kaon pairs at low \q\ ($<0.6$~GeV) 
with at least one kaon from a short-lived source is about 81$\%$, so that 
the fraction of pairs in which both kaons arise from short-lived sources is  
about 66$\%$. 

The pairs from short-lived sources cannot be identified in the 
data, and so the final results of this analysis 
were not corrected for the effect of short-lived sources 
because such a correction would be based on a Monte Carlo model 
with its inherent uncertainties. However, to illustrate 
the magnitude of the effect, a correction was applied 
to the correlation function using the estimated 
fraction of short-lived sources (66$\%$), assuming that kaons
from long-lived sources do not contribute to the correlations. The fitted 
parameters of the corrected correlation function 
are: \lamb\ = 1.27 $\pm$ 0.31 and \Radius\ = 0.55 $\pm$ 0.07 fm, where 
the errors are statistical only. 
\begin{table}[h]
\centering
{\begin{tabular}{|l|l|c|}
\hline 
\hline
                         &origin of kaons    & fraction of kaons in \\
                         &                   & pairs with \qltsix \\
\hline 
long-lived sources       &b, c hadron decays   &12$\%$\\         
life-time $c\tau >$ 10 fm&$\phi$(1020)         & 7$\%$\\
\hline
short-lived sources      &string fragmentation   &40$\%$\\      
life-time $c\tau <$ 10 fm&f$_0$(980)             & 1$\%$\\   
                         &${\mathrm K}^*$(892)   &27$\%$\\
                         &other sources          &13$\%$  \\
\hline 
\hline
\end{tabular}}
\caption[]{\label{t:03} Origins of kaons 
in like-sign pairs with low \q\ in the Monte Carlo.}
\end{table}
\subsection{Final-state interactions} \label{s:5.8}
Charged kaons are subject to both the Coulomb and the strong interactions. In
principle, every pair of like-sign kaons from short-lived sources 
should be corrected for these interactions 
in the data (but not in the Monte Carlo, where they were not simulated). 
To apply a correction to all pairs would result in an overestimate of 
the value of the strength parameter~\cite{bowlercoloumb}. 
As in Section~\ref{s:5.7}, the final 
results of this analysis were not corrected for the electromagnetic and strong   
interactions because of the model-dependence of 
such corrections.  
 
As a check of the possible magnitude of any correction, 
the electromagnetic repulsion of like-sign pairs was corrected for
in the data. The like-sign kaon 
pair \q\ spectrum was corrected using the 
Gamow factor~\cite{gamowpaper} $G(\eta) = 2\pi\eta/(e^{2\pi\eta} - 1)$, 
where $\eta = \alpha_{em}\thinspace m_{{\mathrm K}}/Q$, $\alpha_{em}$
is the electromagnetic coupling constant and $m_{{\mathrm K}}$ the
kaon mass. On the assumption that all pairs are from short-lived sources, the fitted
strength and radius of the Coulomb corrected correlation function are:  
\lamb\ = 0.92 $\pm$ 0.25 and \Radius\ = 0.61 $\pm$ 0.17 fm, where 
the errors are statistical only. 
The correlation function was also corrected for both 
the effect of short-lived sources as in Section~\ref{s:5.7} 
and the Coulomb effect, resulting in the 
fitted parameters: \lamb\ = 1.36 $\pm$ 0.55 and \Radius\ = 0.58 $\pm$ 0.11
fm, where the errors are statistical only.

\section{Discussion and conclusions} \label{s:6}

Bose-Einstein correlations have been measured in
identified pairs of charged kaons in hadronic \zzero\ decays 
using the OPAL experiment at LEP. The analysis was 
performed in events with a clear two-jet topology, a requirement which 
was necessary to obtain a suitable reference sample 
using a hemisphere-mixing technique. 
Monte Carlo simulation was used to correct for imperfections in the reference sample by use of a double-ratio for the correlation function. 
The enhancement was parametrised using a Gaussian formula, resulting in a 
strength 
\begin{eqnarray} 
\lamb  & = & \lambresult \nonumber  \nonumber 
\end{eqnarray}and a kaon emitter radius 
\begin{eqnarray} 
\Radius & = & \radresult ~{\mathrm fm}. \nonumber
\end{eqnarray}

A definite conclusion from the present analysis is 
a confirmation of the results of reference~\cite{kkpaperDELPHI}, that 
there are indeed BE correlations in \kakalik\ pairs from 
hadronic \zzero\ decays. 
This implies that there should be such correlations 
in \ksks\ pairs~\cite{preprintalex_lipkin}; therefore it is unlikely 
that the previously observed threshold 
enhancements~\cite{kskspaperALEPH,kskspaperOPAL,kskspaperDELPHI,kkpaperDELPHI}
can be attributed entirely to the f$_0(980)$ decay into kaons.

Values of \lamb\ and \Radius , as measured in hadronic \zzero\ 
decays at LEP for various particle types, are listed for comparison in Table~\ref{t:04}. Since there is evidence~\cite{pipieebarpaperOPAL} that
\lamb\ and \Radius\ may depend on the event topology, the table gives the type of
event used in the various measurements. The reference sample types used in
each analysis are also listed: these may be event- or hemisphere-mixed, 
unlike-sign or Monte Carlo pairs.
\begin{table}[h!]
\centering
{\begin{tabular}{|l|l|l|l|l|l|}
\hline
\hline
 pair 
& ~~~~~~~~~~ \lamb\                        
& ~~~~~~ \Radius\ (fm)
& ref. & events & experiment \\
& ~~~~~~~~~(\textit{stat.}) (\textit{sys.}) 
& ~~~~~~~~~(\textit{stat.}) (\textit{sys.})
& sample 
&        
&      \\
\hline
 \kakalik\ & \lambresulttable\  & \radresulttable\  
& mixed~$\dag$  
& two-jet 
& this analysis \\
           & 0.82 $\pm$ 0.11 $\pm$ 0.25 & 0.48 $\pm$ 0.04 $\pm$ 0.07 
& unlike 
& all 
& DELPHI~\cite{kkpaperDELPHI} \\
\hline
\ksks\   & 1.14 $\pm$ 0.23 $\pm$ 0.32 & 0.76 $\pm$ 0.10 $\pm$ 0.11 
& MC
& all 
& OPAL~\cite{kskspaperOPAL}\\
        & 0.96 $\pm$ 0.21  $\pm$ 0.40 & 0.65 $\pm$ 0.07 $\pm$ 0.15
& MC
& all 
& ALEPH~\cite{kskspaperALEPH}\\
         & 0.61 $\pm$ 0.16 $\pm$ 0.16 & 0.55 $\pm$ 0.08 $\pm$ 0.12 
& MC
& all  
& DELPHI~\cite{kkpaperDELPHI}\\
\hline
\pipilik\  & 0.67 $\pm$ 0.01 $\pm$ 0.02 & 0.96 $\pm$ 0.01 $\pm$ 0.02 
& unlike
& all 
& OPAL~\cite{pipieebarpaperOPAL}\\
         & 0.65 $\pm$ 0.02 ~~ ----- & 0.91 $\pm$ 0.01 ~~ -----   
& unlike 
& two-jet 
& OPAL~\cite{pipieebarpaperOPAL}\\
         & 0.40 $\pm$ 0.02 ~~ -----~$\ddag$ & 0.49 $\pm$ 0.02 ~~ -----   
& mixed~$\dag$ 
& two-jet 
& ALEPH~\cite{pipieebarpaperALEPH}\\
         & 0.62 $\pm$ 0.04 ~~ -----~$\ddag$ & 0.81 $\pm$ 0.04 ~~ ----- 
& unlike~$\dag$ 
& two-jet 
& ALEPH~\cite{pipieebarpaperALEPH}\\
         & 0.35 $\pm$ 0.04 ~~ -----~$\ddag$ & 0.42 $\pm$ 0.04 ~~ -----               
& mixed~$\dag$ 
& two-jet 
& DELPHI~\cite{pipieebarpaperDELPHI}\\
         & 0.45 $\pm$ 0.02 ~~ -----~$\ddag$ & 0.82 $\pm$ 0.03 ~~ ----- 
& unlike~$\dag$ 
& all 
& DELPHI~\cite{pipieebarpaperDELPHI}\\
\hline 
\hline 
\end{tabular}}
\caption{\label{t:04} Summary of the parameters
\lamb\ and \Radius\ measured at LEP using the Gaussian 
parametrisation and for different types of identical pairs.The results
marked with $\dag$ were obtained using the double-ratio technique for the
correlation function. All 
the \pipilik\ results were
corrected for Coulomb interactions and the ones marked
with $\ddag$ were in addition corrected for non-pion impurities. The uncertainties are statistical and systematic
as labelled.}
\end{table}

In all events and for correlations measured using the 
unlike-sign reference sample, the radius of charged pion 
emitters varies between 0.8 and 1.0~fm while the radius of the 
charged kaon emitters is $0.48 \pm 0.08$~fm. This gives the relation
$R_0(\pipilik) > R_0(\kakalik)$ --- a mass dependence of
the emitting source, as already pointed out  
in~\cite{preprintGideon,preprintBialas}.
In two-jet events, the measured radius of charged pion emitters is seen 
to vary between 0.4 and 0.9~fm, inconsistent within 
the quoted errors. This large variation is usually attributed to the choice
of the reference sample: in the case of the unlike-sign reference
sample the radius is about 0.8--0.9~fm; in the case of 
the mixed reference sample the radius is about 0.4--0.5~fm.
The comparison of results obtained using the event-
or hemisphere-mixing techniques and a double-ratio for the correlation
function shows that the present measurement of the
radius of the kaon source, $0.56 \pm 0.11$~fm, is compatible with the previous 
measurements of the radius of pion sources and does not support a 
strong mass dependence of the emitting source.  
However, both measurements of \Radius\ for kaon pairs are 
considerably larger than that obtained for $\Lambda\Lambda$ pairs,  
0.14$^{+~0.07}_{-~0.03}$~fm~\cite{preprintGideon,lambdapaperOPAL}.
\eject
\bigskip\bigskip\bigskip
\appendix
\par
Acknowledgements:
\hfil\break
\par
We particularly wish to thank the SL Division for the efficient operation
of the LEP accelerator at all energies
 and for their continuing close cooperation with
our experimental group.  We thank our colleagues from CEA, DAPNIA/SPP,
CE-Saclay for their efforts over the years on the time-of-flight and trigger
systems which we continue to use.  In addition to the support staff at our own
institutions we are pleased to acknowledge the  \\
Department of Energy, USA, \\
National Science Foundation, USA, \\
Particle Physics and Astronomy Research Council, UK, \\
Natural Sciences and Engineering Research Council, Canada, \\
Israel Science Foundation, administered by the Israel
Academy of Science and Humanities, \\
Minerva Gesellschaft, \\
Benoziyo Center for High Energy Physics,\\
Japanese Ministry of Education, Science and Culture (the
Monbusho) and a grant under the Monbusho International
Science Research Program,\\
Japanese Society for the Promotion of Science (JSPS),\\
German Israeli Bi-national Science Foundation (GIF), \\
Bundesministerium f\"ur Bildung, Wissenschaft,
Forschung und Technologie, Germany, \\
National Research Council of Canada, \\
Research Corporation, USA,\\
Hungarian Foundation for Scientific Research, OTKA T-029328, 
T023793 and OTKA F-023259.\\



\begin{figure}[htbp]
  \begin{center}
   \epsfysize=14cm
   \epsffile{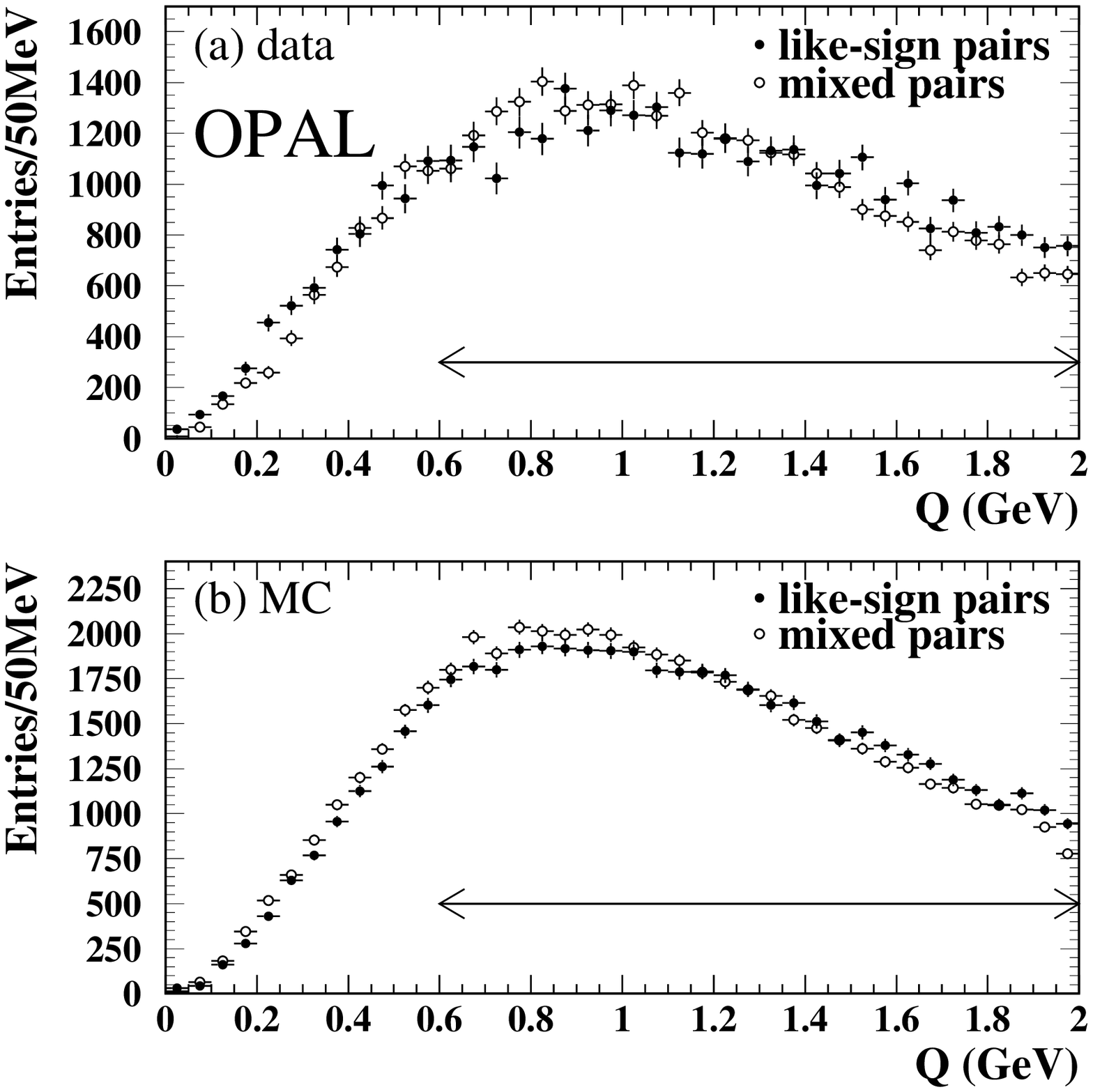}
  \end{center}
\caption[]{\label{f:01}
\q\ distributions of like-sign and hemisphere-mixed \kaka\ pairs in
the data (a) and in the Monte Carlo without simulation of BE
correlations (b). The
distributions were corrected for non-\kaka\ impurities. The 
hemisphere-mixed distributions were normalised to the like-sign
distributions in the region \qnorm , indicated by the arrows, where no 
BE correlations are expected. The errors are statistical only.}
\end{figure}
\begin{figure}[htbp]
  \begin{center}
   \epsfysize=13cm
   \epsffile{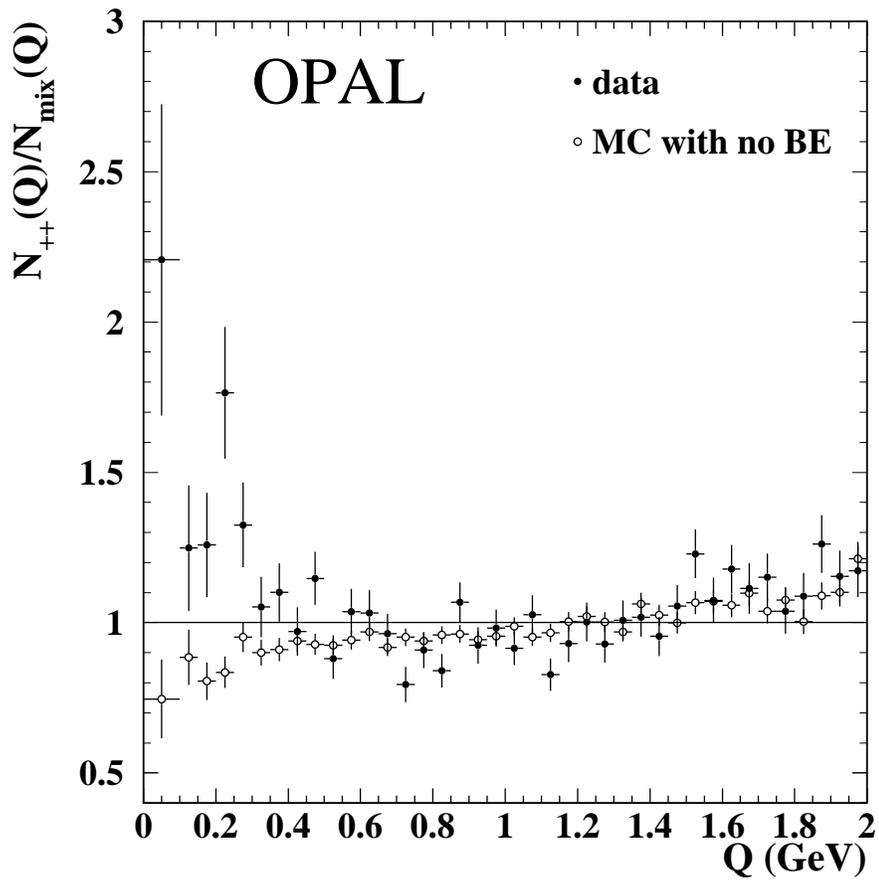}
  \end{center}
\caption[]{\label{f:02}
Ratio $N_{++}(Q)/N_{mix}(Q)$ for \kaka\ pairs in the data
and the Monte Carlo. The first bin has double width due 
to the small statistics. The errors are statistical only.}
\end{figure}
\begin{figure}[htbp]
  \begin{center}
   \epsfysize=13cm
   \epsffile{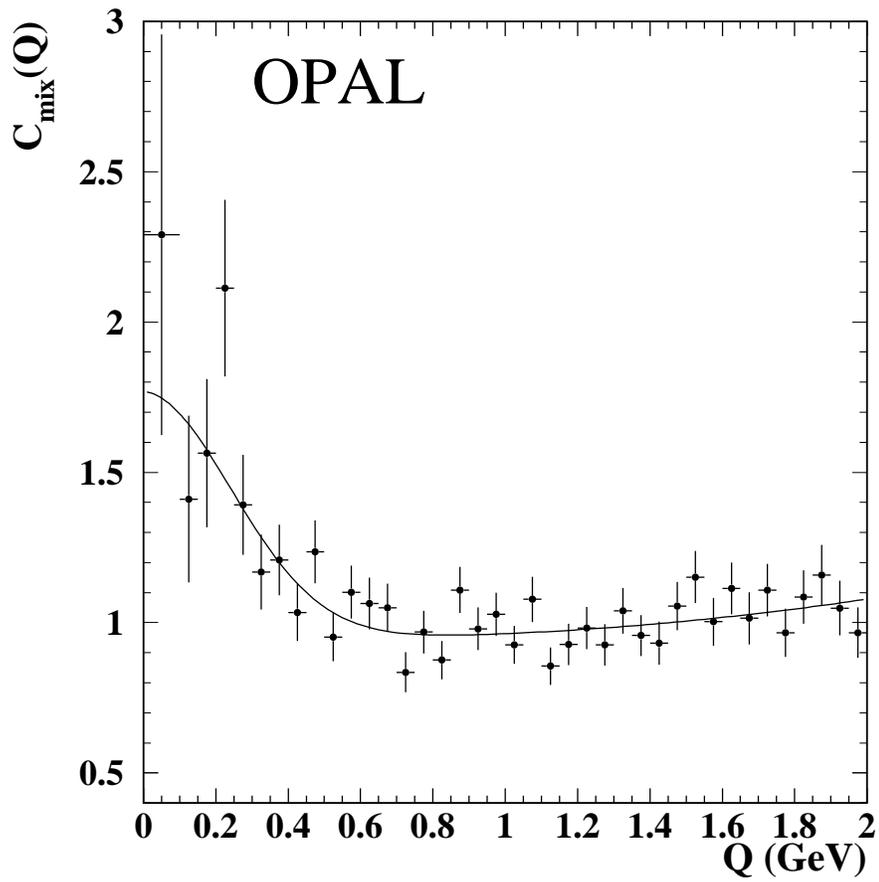}
  \end{center}
\caption[]{\label{f:03}
$C_{mix}(Q)$ BE correlation function in \kakalik\ pairs and 
the fit using Equation~\ref{eq:Cfinal} superimposed. The first bin 
has double width due to the small statistics. The errors are statistical only.}
\end{figure}
\begin{figure}[htbp]
  \begin{center}
   \epsfysize=13cm
   \epsffile{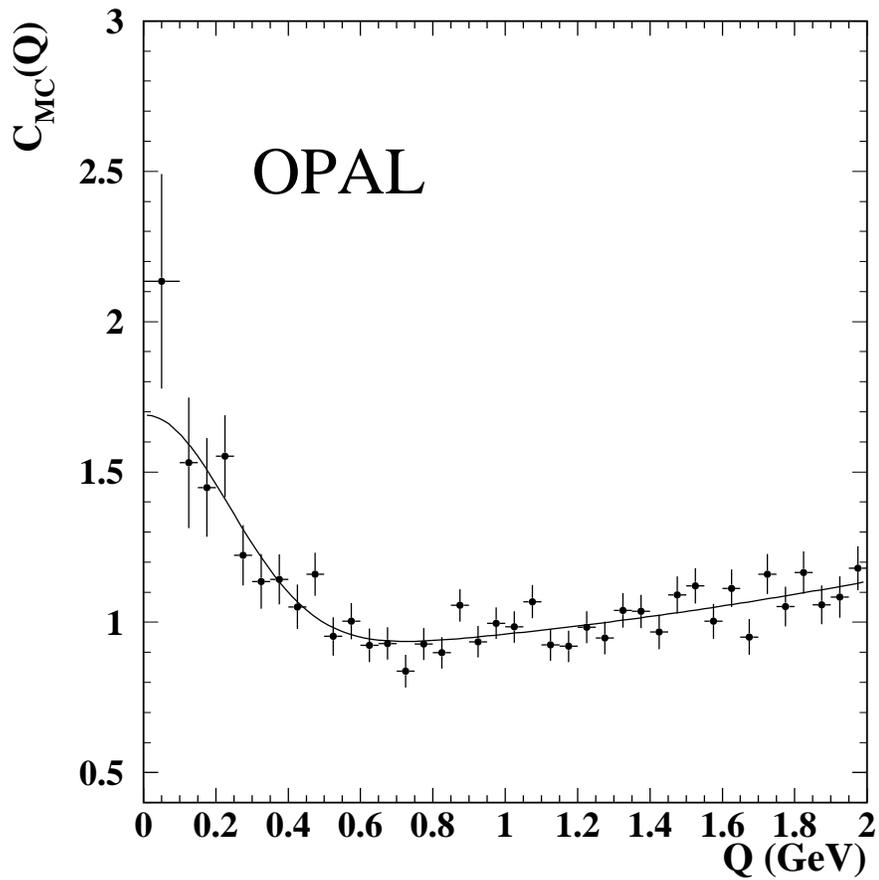}
  \end{center}
\caption[]{\label{f:04}
$C_{MC}(Q)$ BE correlation function in \kakalik\ pairs using 
Monte Carlo like-sign pairs as the reference sample and the fit 
using Equation~\ref{eq:Cfinal} superimposed. The first bin 
has double width due to the small statistics. The errors
are statistical only.}
\end{figure}
\begin{figure}[htpb]
  \begin{center}
   \epsfysize=13cm
   \epsffile{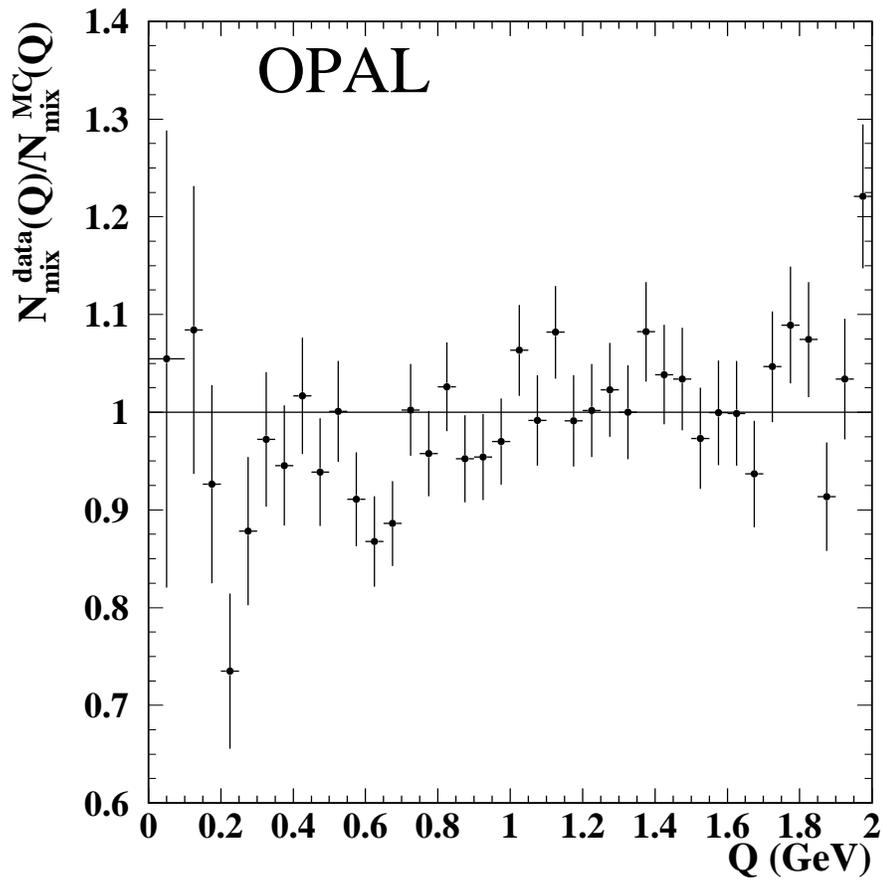}
  \end{center}
\caption[]{\label{f:05}
Comparison of the hemisphere-mixed \q\ distributions in the data and
the Monte Carlo using the ratio $N_{mix}^{data}(Q)/N_{mix}^{MC}(Q)$. The 
errors are statistical only.}
\end{figure}
\begin{figure}[htbp]
  \begin{center}
   \epsfysize=13cm
   \epsffile{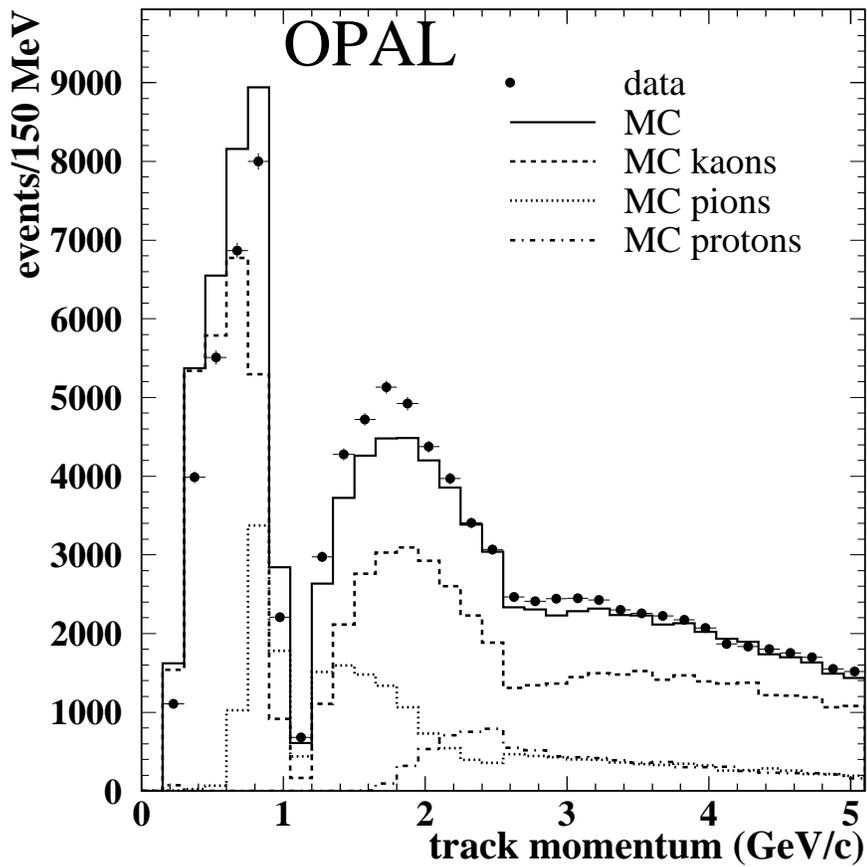}
  \end{center}
\caption[]{\label{f:06}
Momentum distribution of kaon candidates in the data
and the Monte Carlo. The spectra for true kaons, pions and protons in
the Monte Carlo are also shown.}
\end{figure}
\end{document}